\documentclass[11pt,a4paper]{article}
\usepackage{jcappub}

\usepackage{graphicx}

\newcommand{\beq}{\begin{equation}}
\newcommand{\eeq}{\end{equation}}
\newcommand{\beqa}{\begin{eqnarray}}
\newcommand{\eeqa}{\end{eqnarray}}
\newcommand{\beqar}{\begin{eqnarray*}}
\newcommand{\eeqar}{\end{eqnarray*}}

\begin{document}




\title{Propagation of $B$ mesons in the atmosphere}
\author{A.~Bueno, A.~Gasc\'on, J.I.~Illana, M.~Masip}
\affiliation{
CAFPE and Departamento de F{\'\i}sica Te\'orica y del Cosmos \\
Universidad de Granada, E-18071, Granada, Spain}

\emailAdd{a.bueno@ugr.es}
\emailAdd{agascon@ugr.es}
\emailAdd{jillana@ugr.es}
\emailAdd{masip@ugr.es}

\abstract{
Collisions of cosmic rays in the atmosphere may produce 
heavy hadrons of very high energy. The decay length of a 
$B$ meson of energy above $10^7$~GeV is larger than 1 km, 
implying that such a particle tends to interact in the air 
before it decays. We 
show that the fraction of energy deposited
in these interactions is much smaller than in proton and pion
collisions. We parameterize their elasticity 
and determine the average number of interactions and the atmospheric
depth at the decay point for different initial energies. 
We find that the profile of a $3\times 10^9$~GeV bottom shower 
may be very different from the profile
of a proton shower of the same energy, defining either a very
deep maximum, or two maxima, or other features that 
cannot be parameterized with a single Gaisser-Hillas function.
Finally, we discuss under what conditions a bottom hadron
inside the parent air shower may provide
observable effects.}

\keywords{bottom quark, hadrons, cosmic rays, propagation}

\arxivnumber{1109.4337}

\maketitle

\section{Introduction}
Cosmic rays produce collisions in 
the upper atmosphere of energy well above the ones 
observed in particle colliders. Our knowledge of the hadron
properties must then be extrapolated to such energies,
a procedure that introduces some important uncertainties. 
These uncertainties are not only related to the appearance
of new particles and interactions, but also to the 
{\it transition} to a regime where the properties of the standard
particles could be substantially different. In particular,
heavy hadrons that decay through weak interactions 
have typical decay lengths
$\lambda^{\rm dec}_0=0.1$--$0.5$ mm. 
As their energy grows, Lorentz dilation
will eventually make $\lambda^{\rm dec}_0$ longer than the 
interaction length in the air, implying that they tend to 
collide before decaying. Obviously, we do not
know how a charmed or a bottom hadron behaves inside a calorimeter,  
since in colliders they decay 
well before they can reach it. Although heavy-hadron 
interactions are of no interest there,
they are processes that might occur in extensive air showers
and could introduce unexpected effects. 
The production of standard heavy-hadrons 
in the atmosphere has been extensively studied in the
literature (see \cite{Costa:2000jw} for a review), as
its prompt decay is an important source of muons
of very high energy 
(see also \cite{Illana:2009qv,Illana:2010gh}), however,
their propagation has been basically ignored.

In a high-energy collision with a target at rest,
a proton or a pion will break
into several pieces, with each piece 
taking a similar amount of energy. The elasticity $z$
(fraction of energy taken by the leading hadron) and
the multiplicity in such collisions will determine how 
the hadronic shower develops through the atmosphere, which
is equivalent to 10 meters of water if crossed vertically. 
It is clear, however, that if the 
collision involves a charm or, especially, a bottom hadron 
the situation will be very different, as the 
piece carrying the heavy quark will take most of the initial energy
after the collision. Therefore, qualitatively one expects
lower inelasticities and, as
a consequence, a different profile of deposited energy.

In this paper we will explore the propagation of 
$b$ hadrons in the atmosphere. 
In Section 2
we will follow the method applied for charmed 
hadrons in \cite{Barcelo:2010xp} to parameterize the 
elasticity of their collisions with air nuclei. 
In Section 3 we use {\sc CORSIKA} \cite{Heck:1998vt}
to find the profile of bottom showers
of different energies.
In Section 4 we discuss under what circumstances a bottom
component inside an extensive air shower
may give an observable effect. Section 5 is devoted to
the summary and discussion.

\section{Inelasticity in heavy-hadron collisions}
In this section we will modify {\sc PYTHIA} \cite{Sjostrand:2006za} 
to simulate the collisions of a hadron
$H$ (containing a $b$ quark, $H=\Lambda_b,B$) with matter. 
{\sc PYTHIA} describes hadronic collisions of pions and protons,  
distinguishing two types of interactions: {\it diffractive}
processes, where the two hadrons (as a whole) exchange momentum 
through pomerons, and {\it non-diffractive} or {\it partonic} ones,
where gluons are exchanged between the partons in
the colliding hadrons. The latter type includes both soft
collisions of $q^2\le 1$~GeV and the hard ones in
deep inelastic scatterings. 

In {\sc PYTHIA} non-diffractive 
processes dominate the inelastic 
cross section. For example, the cross section 
of a $10^9$~GeV proton or pion with a proton at rest is
\beq
\sigma^{\textrm{\scriptsize{pp}}}=94.1\; {\rm mb}\,; \;\;\; 
\sigma^{\pi\textrm{\scriptsize{p}}}=64.8\; {\rm mb}\,, 
\eeq
where
diffractive processes contribute just 30\% in pp
collisions and 26\% in $\pi$p interactions. Therefore,
we will study diffractive and partonic processes separately, 
then we will combine them to define a generic $H$p scattering,
and finally we will discuss the collision with an air nucleus.

\subsection{Diffractive \texorpdfstring{$H$}{}p collisions}

Let us first describe the approach
to simulate a diffractive $H$p
collision with {\sc PYTHIA} from the analogous light-hadron process.
A hadron $H$ with mass m$_{H}$ contains a heavy core of mass
$m_b\approx 4.7$~GeV. One may think of
a proton as three  clouds of mass 
$m=0.3$~GeV associated to the three constituent quarks. 
Within that picture, a $\Lambda_b$ baryon will consist also 
of three similar clouds,  
but with one of them having an additional electroweak core 
and a total mass $m_b+m=5.0$~GeV.
In a diffractive 
process, however, this heavy core
will be  invisible (the proton and the $\Lambda_b$ clouds 
will look identical), since the interaction has a
$q^2$ much smaller than 1~GeV$^2$ and 
cannot resolve it. Therefore, the momentum
exchanged through pomerons or other non-perturbative dynamics with
the target should not depend on the electroweak core in $H$.

More precisely, the light degrees of freedom in $H$ carry just
a fraction $w\equiv (m_H-m_b)/m_H$ of the hadron energy $E$.
In a diffractive scattering, $H$ will be  seen by the target
nucleon like a light hadron of energy $wE$. Therefore,
to estimate the momentum $q^\mu$ absorbed by $H$ 
 we will just simulate with {\sc PYTHIA}
the collision of a proton (for $H=\Lambda_b$) or a pion
(for $H=B$) of energy $wE$ with the nucleon and
assume that $q^\mu$ is the same when the incident particle
is the heavy hadron.
 
This momentum is all we need to simulate how $H$
evolves after the diffractive collision 
(see \cite{Barcelo:2010xp}  for details). 
Once $H$ absorbs $q^\mu$, it becomes a system 
of mass $M^*$, the critical
parameter in the collision. If $M^*< m_H+1$~GeV the process is
 quasi-elastic and $H_{di\! f}$ will just decay into two
bodies (e.g., $H+\eta$). For larger values of $M^*$ the system
is treated by {\sc PYTHIA} \cite{Navin:2010kk}
like a string with the quantum numbers of
$H$. When $H$ is a baryon the string may be stretched 
between a quark and a diquark or between a 
quark, a gluon and a diquark, whereas for a diffractive meson
the string connects a quark and an antiquark or a quark, a gluon 
and an antiquark. 

\begin{figure}[!t]
\begin{center}
\includegraphics{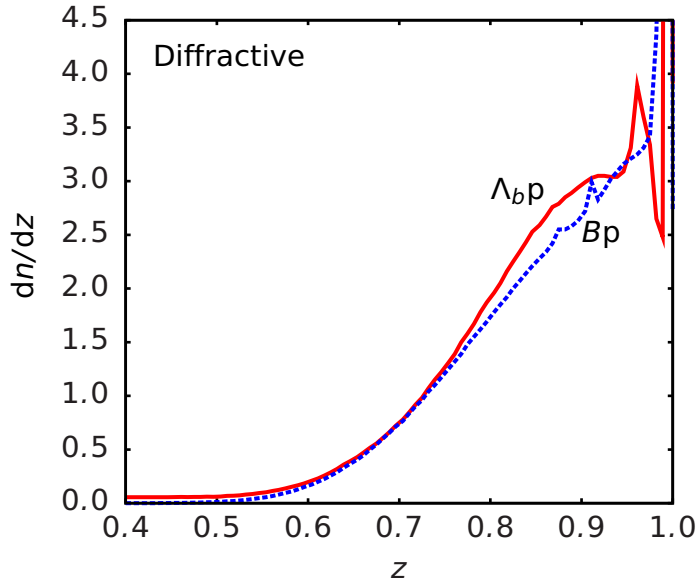} 
\end{center}
\caption{Elasticity ($z$) in $\Lambda_b$p (solid) and $B$p (dashed)
diffractive collisions.
\label{fig1}}
\end{figure}

In Fig.~\ref{fig1} we plot the distribution of the fraction $z$ of 
energy taken by the bottom hadron after a diffractive 
collision of a $10^9$~GeV $\Lambda_b^0$ or a 
$B$ with a proton at rest.
The average values are $\langle
z\rangle =0.88$ and $\langle z\rangle =0.90$, respectively, which
imply an inelasticity $K=1-\langle z\rangle$:
\beq 
K_{\Lambda_b\textrm{\scriptsize{p}}}^{di\!
f}=0.12\;;\;\;\;\;K_{B\textrm{\scriptsize{p}}}^{di\! f}=0.10\,. 
\eeq 
In comparison, we note that for proton and charged-pion collisions, 
the inelasticity obtained also with {\sc PYTHIA} 
is substantially higher:
\beq 
K_{\textrm{\scriptsize{pp}}}^{di\!
f}=0.40\;;\;\;\;\;K_{\pi\textrm{\scriptsize{p}}}^{di\! f}=0.23\,. 
\eeq 

\subsection{Partonic \texorpdfstring{$H$}{}p collisions}
For non-diffractive collisions,
we model  $H$
as a system with the same parton content as the
corresponding proton or pion, but substituting a valence up quark
($u_0$) for a bottom quark. As in diffractive processes, we will
associate a hadron $H$ of energy $E$ to a light hadron of
energy $(m_H-m_b)/m_H\,E$. If $u_0$ carries a fraction $x$ of the
proton or the pion momentum, we will change it for a $b$ 
with momentum fraction $x_{b}$
\beq
x_b=\frac{m_b}{m_H} + \frac{m_H-m_b}{m_H} x\,.
\eeq
In this way the
excess of energy in $H$ is carried entirely by the bottom quark,
whereas the light partons in both hadrons ($H$ and p or $\pi$)
carry exactly the same amount of energy.

We will then distinguish two types of $H$p partonic
collisions: those where the bottom is an spectator (i.e., it
is a light parton in $H$ which hits a parton in the target proton),
and processes where the $b$ quark itself interacts. For the first case
we first just simulate with {\sc PYTHIA} the parton process using a light
hadron, later we then substitute the spectator $u_0$  for the
bottom quark. Intrinsic bottom interactions, on the other hand, have a much
smaller cross section, 
\beq 
\sigma_{b\;int}^{H\textrm{\scriptsize{p}}}=0.8 \; {\rm mb}\,, 
\eeq 
than the processes with an spectator $b$ quark, 
\beq
\sigma_{b\;spec}^{\Lambda_b\textrm{\scriptsize{p}}}=56.2 \;{\rm mb}\,; \;\;\;
\sigma_{b\; spec}^{B\textrm{\scriptsize{p}}}=38.7\; {\rm mb}\,, 
\eeq 
but they imply
collisions of higher inelasticity. 

\begin{figure}[!t]
\begin{center}
\includegraphics{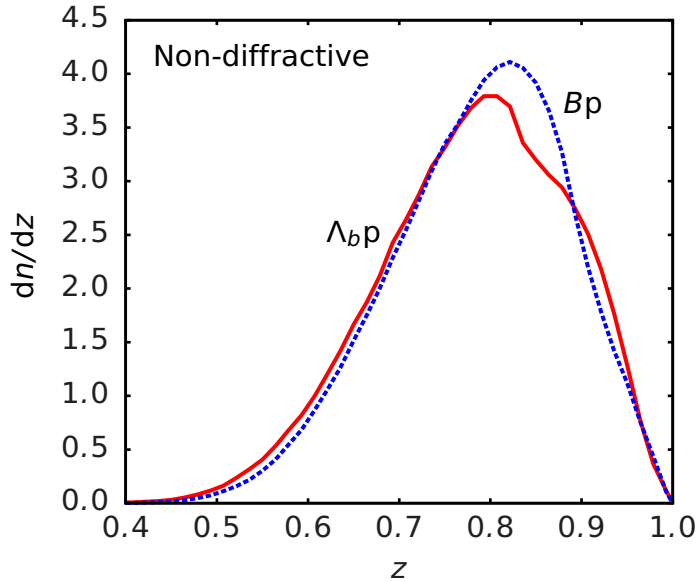} 
\end{center}
\caption{Elasticity ($z$)
in $\Lambda_b$p (solid) and $B$p (dashed) non-diffractive collisions.}
\label{fig2}
\end{figure}

In Fig.~\ref{fig2} we plot the distribution of energy taken by the
bottom hadron after a $\Lambda_b$p or a 
$B$p non-diffractive collision.
The average inelasticity is 
\beq 
K_{\Lambda_b\textrm{\scriptsize{p}}}^{n-di\!
f}=0.22\;;\;\;\;\;K_{B\textrm{\scriptsize{p}}}^{n-di\! f}=0.21\,, 
\eeq 
For comparison for protons and 
 pions of the same energy is:
\beq 
K_{\textrm{\scriptsize{pp}}}^{n-di\!
f}=0.66\;;\;\;\;\;K_{\pi p}^{n-di\! f}=0.77\,. 
\eeq 

\subsection{Inelasticity in \texorpdfstring{$H$}{}p collisions}
We need to combine both types of collisions. 
We find that the total (inelastic) $H$p cross section for
a projectile of energy $10^9$~GeV is 
\beqa 
\sigma^{\Lambda_b\textrm{\scriptsize{p}}}&=&\sigma^{\Lambda_b\textrm{\scriptsize{p}}}_{di\! f}+\sigma^{\Lambda_b\textrm{\scriptsize{p}}}_{n-di\! f}=82.1\;{\rm mb}\,;\nonumber\\ 
\sigma^{B\textrm{\scriptsize{p}}}&=&\sigma^{B\textrm{\scriptsize{p}}}_{di\! f}+\sigma^{B\textrm{\scriptsize{p}}}_{n-di\! f}=54.4\; {\rm mb}\,, 
\eeqa
where
diffractive processes contribute 30\% in $\Lambda_b$p
collisions and 28\% in $B$p interactions. 
The cross sections for
proton and pion collisions 
of the same energy are 14\% and 19\% larger, respectively.

\begin{figure}[!t]
\begin{center}
\includegraphics{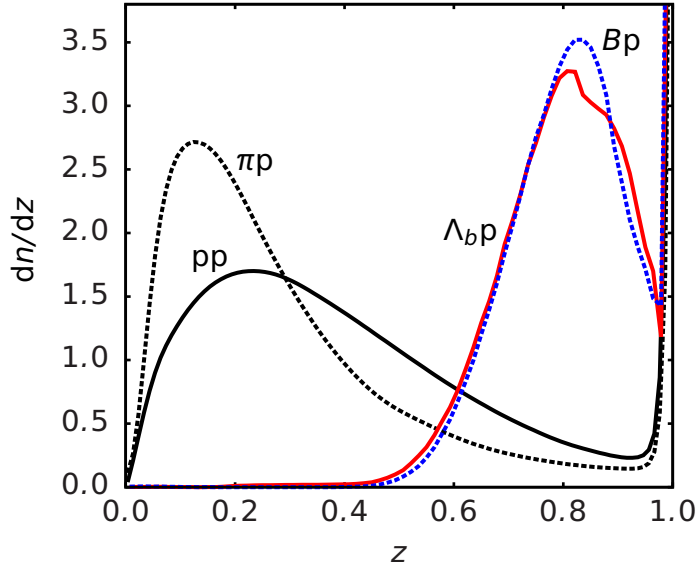} 
\end{center}
\caption{Elasticity ($z$) 
in $(\Lambda_b,B)$--p and 
$(p,\pi)$--p collisions.
\label{fig3}}
\end{figure}

Using the relative frequency of these processes to find
the average inelasticity, we obtain (see Fig.~\ref{fig3}) 
\beq 
K_{\Lambda_b\textrm{\scriptsize{p}}}=0.19\,;\;\;\;\;
K_{B\textrm{\scriptsize{p}}}=0.18\,, 
\eeq 
which is substantially lower than the inelasticity
in proton and pion collisions, 
\beq 
K_{\textrm{\scriptsize{pp}}}=0.59\,;\;\;\;\;K_{\pi\textrm{\scriptsize{p}}}=0.70\,. 
\eeq

\subsection{\texorpdfstring{$H$}{}--air collisions}
\label{ssec:air}
The total
cross section for the collision of $H$ with an atomic nucleus of
mass number $A$ can be approximated as 
\beq 
\sigma^{H A}\approx
A^{2/3} \sigma^{H\textrm{\scriptsize{p}}}\,, 
\eeq 
where the factor of $ A^{2/3}$ takes
into account the screening between the nucleons inside the
nucleus. For an averaged atmospheric nucleus of $A=14.6$ this
implies 
\beq 
\sigma^{\Lambda_b\, air}=490\; {\rm mb}\,; \;\;\;
\sigma^{B\, air}=325\; {\rm mb}\,. 
\eeq 
The associated interaction
length $\lambda^{H}_{int}=m_{air}/\sigma^{H\,air}$ in the
atmosphere is therefore
\beq 
\lambda^{\Lambda_b}_{int}=49\; {\rm g \ cm^{-2}}\,; \;\;\;
\lambda^{B}_{int}=74\; {\rm g \ cm^{-2}}\,, 
\eeq 
which is a 14\% and a 19\% longer
than those of a pion and a proton of the same energy,
respectively.

To deduce the energy and the species
of the $b$ hadron after the collision
we follow \cite{Barcelo:2010xp} and 
distinguish between {\it peripheral} and {\it central}
collisions. We assume that the spectrum in peripheral processes
coincides with the one in $H$p collisions, whereas the
average inelasticity in central processes is the typical in a
partonic scattering increased by $10\%$. In addition, we will take
equal frequency for both types of processes. The average inelasticity
can then be estimated as
\beq
K_{H\,air}\approx frac{1}{2}\, K_{H\, air}^{peri}+ \frac{1}{2}\,
K_{H\, air}^{cent} \approx \left( \frac{1}{2}\, K_{H\textrm{\scriptsize{p}}} + \frac{1}{2}\, 1.1 \, K_{H\textrm{\scriptsize{p}}}^{n-di\! f}\right) \,. 
\eeq 
At $E=10^9$~GeV, we obtain 
\beq 
K_{\Lambda_b\, air}\approx 0.21\,;\;\;\; K_{B\,
air}\approx 0.20\,. 
\eeq 
For proton and pion collisions the same
prescription gives an inelasticity 
\beq 
K _{\textrm{\scriptsize{p}}\, air}\approx
0.66\,;\;\;\; K_{\pi\, air}\approx 0.78\,.
\eeq 
This $12\%$ increase in $K$ when going from a proton to a
nucleus target compares well with the results obtained by 
other authors \cite{Ostapchenko:2009zz}. 

\begin{figure}[!t]
\begin{center}
\includegraphics{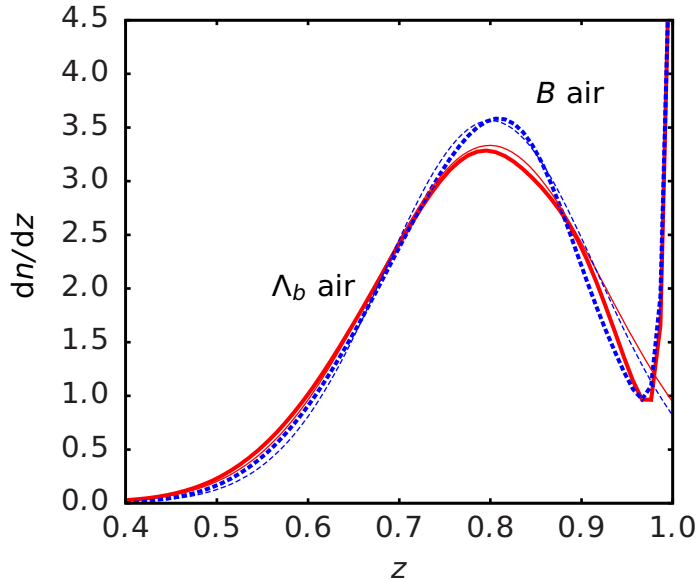} 
\end{center}
\caption{Elasticity ($z$) in 
$\Lambda_b$--air (thick solid) and $B$--air (thick dashed) 
collisions. We include in both cases the fit with the
exponential in Eq. \ref{eq:param}.
\label{fig4}}
\end{figure}

In Fig.~\ref{fig4} we plot (thick lines) the final distribution $z$ of 
the fraction of energy taken
by the heavy hadron after the collision of a $\Lambda_b$ 
or a $B$ 
with an air nucleus. We also plot (thin lines) our parameterization
of the distribution as an exponential between 0 and 1:
\beq
f(z)=\frac{1}{N}e^{ -a(z-z_0)^2}\,,
\label{eq:param}
\eeq
where $N$ normalizes the distribution to 1. We obtain
$a=31.3$, $z_0=0.799$ and $N=0.30$ for an incident $\Lambda_b$ baryon
and $a=37.0$, $z_0=0.800$ and $N=0.28$ for an incident $B$ meson.

\begin{figure}[!t]
\begin{center}
\includegraphics{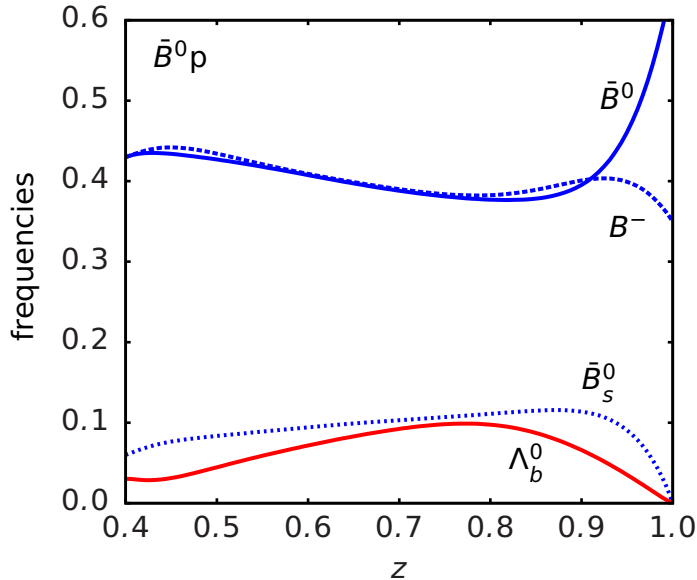} 
\end{center}
\caption{Frequency of the different $b$-hadron species after the 
collision of a $10^9$~GeV $\bar B^0$ meson with a proton at rest.
\label{fig5}}
\end{figure}

Finally, we plot in Fig.~\ref{fig5}
 the frequency of the different hadron species after
the collision of a $10^9$~GeV
$B$ meson with a proton at rest for different 
values of the elasticity $z$. We obtain a $B^0$ $42\%$ of the
times, $B^-$ mesons appear after $41\%$ of the collisions, 
whereas the frequency of $B_s^0$ mesons and $\Lambda_b^0$ baryons is
$9\%$ and $7\%$, respectively. 
In $\Lambda_b$ collisions the approximate frequency of 
($\Lambda_b$, $B^0$, $B^-$, $B_s^0$) 
provided by {\sc PYTHIA} is
($68\%$, $14\%$, $14\%$, $4\%$).

\section{Propagation  in the atmosphere}
\label{sec:propagation}
The propagation of 
a long-lived $b$ hadron differs from that of a proton or a pion, due to its larger elasticity in collisions
with air nuclei and its longer interaction length. Using the 
Monte Carlo code {\sc CORSIKA} we analyze the
energy deposition of a $b$ hadron produced in the upper 
atmosphere, detached from the development of its parent shower.
 We focus on $B^0$ mesons, as $\Lambda_b$
baryons tend to convert into mesons after a few hadronic
interactions. The $B^0$ meson may
interact\footnote{If it deposits an energy 
$\Delta E= (1-z)E$ in 
a collision, we will simulate there a CORSIKA pion shower
of the same energy $\Delta E$.}
(according to the cross section and
the inelasticity described in Section 2)
or decay\footnote{If it decays at a depth $X$ we treat the decay using PYTHIA and
inject the products at that depth using CORSIKA.} as it propagates, producing secondary showers
at different atmospheric depths.

\begin{figure}[!t]
\begin{center}
\begin{tabular}{c}
\includegraphics{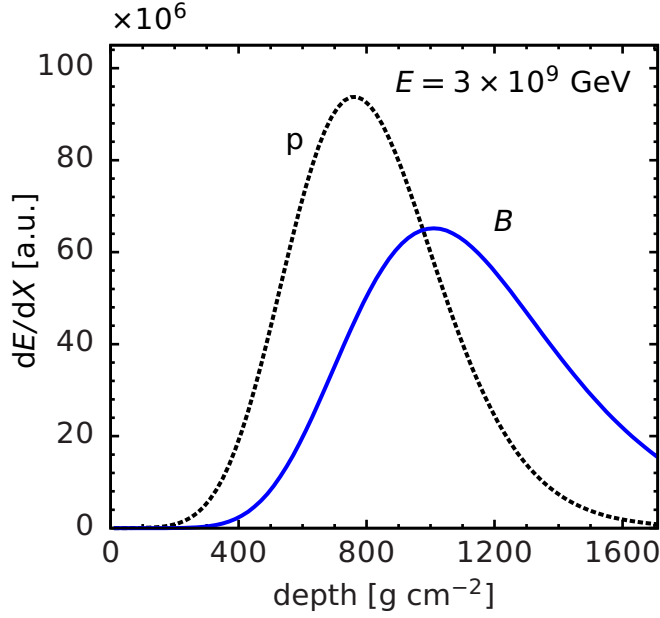} 
\end{tabular}
\end{center}
\caption{Average profile of a p (dashed) and a $B$ (solid) shower for an
initial energy of $3\times 10^9$~GeV.}
\label{fig6}
\end{figure}

As an example, we will take a $B^0$ meson produced
at a depth of $X_0=100$~g cm$^{-2}$ with a zenith inclination 
$\theta=60^\circ$. We consider different 
values of the initial energy of the $B^0$ meson, $E_{B}$. For values
of $E_{B}$ of the order of $10^8$~GeV, the $B^0$ mesons decay after
few interactions and therefore 
look like a proton shower. When larger energies O($10^9$~GeV) are
considered, the differences between $B^0$ and proton initiated showers
become more evident. To stand out the most relevant features of the
showers produced by $B^0$ particles, we consider in this section
average proton shower profiles. The next section, devoted to the detectability of
bottom hadrons produced inside extensive air showers, will deal with
individual fully-simulated showers to take into account the effect
that shower fluctuations introduce in our analysis.

In Fig.~\ref{fig6}  we plot the averaged profile of 100 $B^0$ initiated 
showers (solid) and of the same number of proton showers (dashed). The
initial energy for $B^0$ and protons is $3\times 10^9$~GeV. 
The average $B^0$ decay 
takes place at $\langle X_f \rangle =1190 \pm 340$~g cm$^{-2}$, after $\langle N_I \rangle =16\pm 4$ 
interactions with air nuclei. These $B^0$ showers are 
different from a proton shower of the same energy 
in two important aspects. First, the energy deposition rate in the atmosphere is slower, and 
the shower maximum is reached later, around 1000~g~cm$^{-2}$.
Among the 100 showers generated, 52 reach their maximum beyond 1000~g cm$^{-2}$. 
Also, secondary showers
of energy $10^7$--$10^8$~GeV start deep in the atmosphere
(as a result of the decay or the collisions of the heavy hadron), increasing the number of electromagnetic particles
that reach the ground. 
The profile of these showers, in general, cannot be well
described by a single Gaisser-Hillas function. 

\begin{figure}[!t]
\begin{center}
\includegraphics{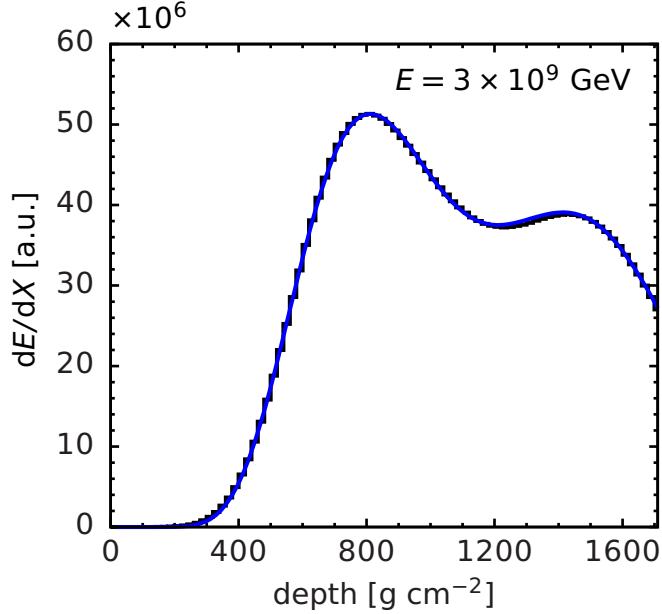}
\end{center}
\caption{Anomalous profile of a $3\times 10^9$~GeV $B^0$ shower (histogram). 
The solid line corresponds to a fit using a combination of two Gaisser-Hillas functions.}
\label{fig7}
\end{figure}

In Fig.~\ref{fig7} we plot as an illustration an anomalous profile found
among these 100 showers, namely a longitudinal development featuring a
double peak topology. We obtain this feature in 15 out of 100 
simulations. In the same figure we plot (solid) a fit 
to this shower using the sum of two Gaisser-Hillas.

\section{Detectability of  \texorpdfstring{$B$}{} mesons inside extensive air showers}
\begin{figure}[!t]
\begin{center}
\includegraphics{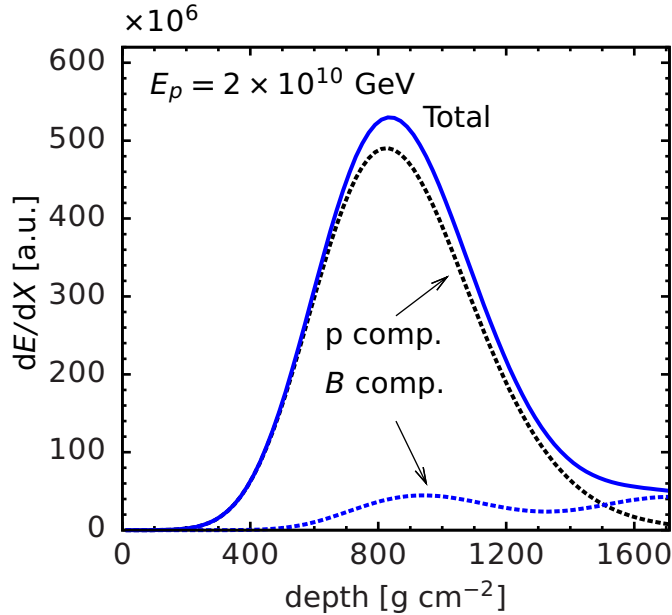} 
\end{center}
\caption{Profile of a $2\times 10^{10}$~GeV proton shower that has produced a
$3\times 10^{9}$ GeV $B$ meson at $X_0=100$~g cm$^{-2}$.
\label{fig8}}
\end{figure}

$B$ mesons are not primary particles
that may start an air shower. They will be  
produced in the collision of a primary cosmic ray 
or an energetic leading 
hadron in the shower with an air nucleus. Therefore, 
the $B^0$ shower described in the previous
section will always be a component 
inside a parent shower. Its observability
will then depend critically on the
fraction $x_b$ of the energy that the $B$ meson 
carries when it is produced. Values $x_b\ge 0.1$ are interesting. 
In Fig.~\ref{fig8}, we plot for the sake of the argument an extreme
example where a {\sc CORSIKA}-simulated proton primary with energy $E_{\textrm{\scriptsize{p}}} =
2\times10^{10}$~GeV and zenith angle $\theta=60^\circ$ 
produces a very energetic $B^0$ meson in the upper atmosphere, 
carrying $E_B=3\times10^{9}$~GeV ($x_b=0.15$).
The rest of the secondary particles produce a shower with energy $E_{\textrm{\scriptsize{p'}}}=
1.7\times10^{10}$~GeV that reaches a maximum around 800 g cm$^{-2}$. 
At ground level the electromagnetic component of this shower has 
been almost completely absorbed.
The $B^0$ meson, produced at $X_0=100$~g cm$^{-2}$ and decaying at
$X=1748$~g cm$^{-2}$ (depth along the proton incoming direction), 
has a deep maximum
and a rich electromagnetic component at the ground. We observe a 
long tail in the shower profile, with $dE/dx$ values
close to the ground that are anomalous in a regular proton shower.

Based on the model described in this work, the simulations show 
that the inclusion of $b$ hadrons in the shower does not produce
a striking deformation of the shower profile, but rather some features 
rarely found in the development of proton showers. Even so, 
these showers will be very often similar to those of protons.
Thus, the search for a bottom component should be done on the basis of 
large statistics, trying to determine which are the observables 
that provide the maximal separation between proton showers
and proton showers with bottom production.

To assess the observability of $b$ quarks inside
inclined extensive air showers, we have produced several samples of
fully simulated extensive air showers. These  {\sc CORSIKA} showers
naturally include fluctuations in the longitudinal development of the
cascades, which can mimic the effects produced by a bottom component. 
All in all,  we have fully simulated several thousands of events for
background and signal componentes:
2000 proton showers for each of the following energies 
log$_{10}$(E/GeV)=\{10, 10.25, 10.5, 10.75\}
and fixed zenith angle of $\theta=60^{\circ}$ stand for our initial
background sample. Per each of the former energies, another 2000 proton showers with a bottom component constitute the signal.

In our simulation, heavy quarks are produced only
in the first interaction of the incoming proton with an air nucleus, 
and not in subsequent interactions
of the leading or secondary 
hadrons with the air: their effect in the shower 
profile will be small due to their smaller energy.
We have taken the cross section for bottom production 
in $p$-air collisions in \cite{Goncalves:2006ch}
(they use a dipole picture in the Color Glass Condensate formalism, 
including saturation effects).
It is found, for example, that at $10^{10}$ GeV this
cross section is around 25 mb, with the bottom component carrying
more than $1\%$ of the initial energy ({\it i.e.} $x_{b\bar b}>0.01$)
in a $25\%$ of the collisions. 
Therefore, in these showers there would be around 1\% probability
to obtain a bottom component of energy above $10^8$ GeV in the
first proton interaction. 
In our analysis we will take a total 
set of $n$ showers, where  
$S_0=0.01\,n$ showers have produced bottom hadrons 
(with the energy distribution in \cite{Goncalves:2006ch}) 
and $B_0=0.99\,n$ are regular proton showers.
The propagation of the bottom component in these showers is performed
according to the results obtained 
in Section \ref{sec:propagation}.

A thorough examination on an event by event basis of the simulated samples revealed a set of 
observables, related to the longitudinal development of the
electromagnetic part of the shower, that help discriminating signal from 
background. The most relevant are:
\begin{itemize}
 \item The logarithm of the number of electromagnetic particles 
reaching the ground.
 \item The difference between the depths beyond $X_{max}$ at which 
the shower is at 40\% and 35\% of the shower maximum.
 \item The difference between the depths before $X_{max}$ at which 
the shower is at 70\% and 75\% of the shower maximum.
 \item The full width at 5\% of the profile maximum.
 \item The central moments of 3$^{\textrm{\scriptsize{rd}}}$ 
and 5$^{\textrm{\scriptsize{th}}}$ order of the shower profile.
 \item The depth beyond $X_{max}$ at which the curvature of 
the Gaisser-Hillas fit to the profile changes.	
\end{itemize}
To enhance the separation power between signal and background,  we
trained a Fisher discriminant using the former observables \cite{Fisher}. 
Applying this discriminant and 
maximizing the value of $S/\sqrt{S+B}$, 
where $S$ and $B$ are the number of selected signal and background events,
we obtain a signal selection efficiency $E^{sel}_S=0.165$ and
a background rejection efficiency $E^{rej}_B=0.992$. Thus, the 
number of selected signal and background events would be
$S=0.00165\,n$ and $B=0.00792\,n$, respectively. This means 
that, according to our model, to get a 3$\sigma$ evidence for production of
$b$ hadrons in extensive air showers (with energy above 
$\approx 10^{10}$~GeV and zenith angle between 55$^\circ$ and 65$^\circ$)
would require to collect around $3\times10^4$ extensive air showers.

\section{Summary and discussion}
Hadrons containing a $b$ quark become long-lived at energies above
$10^7$~GeV, and they tend to collide in the air before decaying. 
We have 
shown that the inelasticity in those collisions is almost four
times smaller than in light-hadron scatterings. 
As a consequence, these hadrons 
can cross a large fraction of atmosphere
keeping a significant fraction of energy. 
For example, a $3\times 10^9$~GeV $B$ meson has an average of 16 interactions
before decaying at $X_f\approx 1200$~g cm$^{-2}$. The final energy of 
the meson is around $10^{7.5}$~GeV, and it will be all deposited close
to the ground. We find that the profile of these showers frequently presents
a very deep shower maximum and occasionally a double maximum.

Although $b$ hadrons are rarely produced in extensive air showers, we have also discussed whether 
they can provide a significant deviation once included 
in the shower development. We have shown that if the heavy meson
decay occurs deep enough in the atmosphere the profile could 
reveal such a feature. A multivariate analysis of
observables defined on the shower electromagnetic profile shows that,
according to our model, a very large number of events need
to be collected in order to get an evidence of $b$ hadron production
in ultra-high energy cosmic rays. 


Despite this result, we think that the production and propagation 
of heavy hadrons should be included in Monte Carlo codes like 
{\sc AIRES} \cite{sergio} or {\sc CORSIKA} \cite{Heck:1998vt}, 
which simulate extensive air showers 
of energy up to $10^{11}$~GeV. Such an effort seems necessary
to establish on a solid ground the expected frequency of events with late
energy deposition and the best 
strategy  in the search for observable effects.

\section*{Acknowledgments}
This work has been partially supported by
MICINN of Spain (FPA2010-16802, FPA2006-05294,   
FPA2009-07187, Consolider-Ingenio 
Multidark CSD2009-00064)  
and by Junta de Andaluc\'{\i}a
(FQM~101, FQM~330, FQM~437 and FQM~3048).

\end{document}